\documentclass[conference,a4paper]{IEEEtran}
\IEEEoverridecommandlockouts

\usepackage{cite}
\usepackage{amsmath,amssymb,amsfonts}
\usepackage{graphicx}
\usepackage{textcomp}
\usepackage{xcolor}
\usepackage{booktabs}
\usepackage{url}

\begin{document}

\title{ChainWatch: A Kill Chain-Aligned Sequential Detection Framework
for Multi-Step Attacks in MCP-Based AI Agent Systems}

\author
{\IEEEauthorblockN{1\textsuperscript{st} Om Narayan}
\IEEEauthorblockA{\textit{Computer Science} \\
\textit{New York University}\\
New York, USA \\
on371@nyu.edu}
\and
\IEEEauthorblockN{2\textsuperscript{nd} Rashmi  Jyoti}
\IEEEauthorblockA{\textit{Cybersecurity} \\
\textit{University of Maryland, College Park}\\
MD, USA \\
rjyoti@umd.edu}
\and
\IEEEauthorblockN{3\textsuperscript{rd} Ramkinker Singh}
\IEEEauthorblockA{
\textit{Carnegie Mellon University}\\
Pittsburgh, USA \\
ramkinks@alumni.cmu.edu}

\thanks{This work is not related to the author's position at Amazon Web Services.}
}

\maketitle

\begin{abstract}
The Model Context Protocol is an open source standard that allows AI
agents to connect to external tools, databases, and services. While this
connection is seamless, it also enables a class of multi-step attacks
that existing per-call defenses have a gap in their mechanism to detect.
Against undefended systems, these attacks achieve over 90\% success rates
by composing individually innocent tool calls into malicious
sequences~\cite{b5}. Existing MCP defenses evaluate each invocation in
isolation and have no mechanism to detect intent that only becomes
visible across a session. This paper presents ChainWatch, a sequential
detection framework designed for this threat class. ChainWatch models
attack progression as a six-stage MCP kill chain, classifies tool-call
sequences using a Hidden Markov Model, and fires detection rules when a
session pattern matches a suspicious progression pattern. The framework
is grounded in a structured threat analysis covering three multi-step
attack categories --- Direct Sequential Attacks, Indirect Injection
Chains, and Hybrid Multi-Stage Attacks. It operationalises these through
a 20-dimensional feature extraction schema derived from documented attack
behaviour. We trace the framework through five scenarios drawn from the
security research literature, demonstrating how ChainWatch would detect
attack chains that pass per-call inspection by existing MCP defenses.
\end{abstract}

\begin{IEEEkeywords}
Model Context Protocol; MCP security; multi-step attack detection;
kill chain; Hidden Markov Model; agentic AI security; LLM agent security
\end{IEEEkeywords}

\section{Introduction}

Before the Model Context Protocol, every AI agent framework built its
own tool integration layer. MCP, released by Anthropic in November 2024,
replaced that fragmentation with a single open standard. Within eighteen
months it had thousands of active servers, integration into Cursor,
Claude Desktop, and GitHub Copilot, and formal stewardship under the
Linux Foundation~\cite{b1,b2}.

The security situation has not kept pace. Over 50 CVEs have been
documented; nearly half of deployed servers have authentication problems,
and researchers have demonstrated real exploits across financial
services, developer tools, and enterprise
messaging~\cite{b2,b3,b4}.

The attack pattern that concerns us most is the kind that existing
defenses cannot see. An attacker who wants to steal credentials does not
need to make a single obviously malicious call. They can check what
tools are available, establish a normal-looking pattern, retrieve
environment variables, read SSH configuration, and post everything to a
webhook --- each step legitimate in isolation; the sequence clearly
malicious. STAC demonstrated that this kind of chained attack achieves
over 90\% success against GPT-4.1~\cite{b5}. The Promptware Kill Chain
analysis found that 21 of 36 documented AI agent attacks traverse four
or more stages~\cite{b6}.

Recent MCP defenses --- MCPShield~\cite{b7}, MCP-Guard~\cite{b8}, and
MindGuard~\cite{b9} --- address per-call and per-server threats
effectively. None of them model whether a session as a whole is following
an attack pattern. Section~\ref{sec:bg} examines each system in detail
and explains why this gap exists by design.

ChainWatch is designed to address that gap. It sits between the MCP
client and its servers as a transparent proxy, observes every tool call
and response, and detects when the sequence of calls across a session
follows a multi-step attack pattern. This paper makes four contributions.
We first develop a threat analysis for multi-step MCP attacks, comprising
a structured attacker model and a classification of three attack types
--- Direct Sequential Attacks, Indirect Injection Chains, and Hybrid
Multi-Stage Attacks. From this threat analysis, we derive a six-stage
MCP Kill Chain that maps observable tool-call features to attack
progression stages. We then present ChainWatch, a sequential detection
framework that operationalises the kill chain through a 20-dimensional
feature extraction schema, an HMM-based stage classifier, and five
session-level detection rules. Finally, we trace the framework through
five scenarios from the security research literature, demonstrating how
ChainWatch would detect each attack class.

\section{Background and Related Work}
\label{sec:bg}

\subsection{MCP Architecture}

MCP defines a three-component architecture over JSON-RPC
2.0~\cite{b1}. The \emph{Host} is the user-facing AI application ---
Claude Desktop, Cursor, or a custom agent. The \emph{Client} manages
connections to MCP servers on the host's behalf. The \emph{Server}
exposes tools, resources, and prompt templates. When a user makes a
request, the host uses the client to query available tools, the model
reads their descriptions to decide which to invoke, and the client
executes the call against the server and returns the result.

Tools are the primary attack surface because the model treats their
descriptions as trusted input when planning which actions to take. Two
properties of the MCP specification are directly relevant to this paper.
First, tool definitions can be changed server-side after the user has
approved them --- the client has no built-in mechanism to detect or alert
on this change. Second, the specification requires no audit logging of
tool invocations~\cite{b1}. The first property enables definition-based
attacks including rug-pulls. The second means sequential attack patterns
leave no protocol-level trace for defenses to analyse.

\subsection{MCP Attack Landscape}

MCP attacks span a spectrum from single-call exploits to coordinated
multi-step chains. Single-call attacks are well documented. Multi-step
attacks --- where no individual call is suspicious but the sequence is
malicious --- are the underdefended class this paper addresses.

Tool poisoning embeds malicious instructions inside tool
descriptions~\cite{b3}. CyberArk Labs extended this to tool outputs ---
injecting instructions through error messages and execution
results~\cite{b22}. MCPTox evaluated these attacks against 45 real-world
MCP servers and found 72.8\% success rates against o1-mini~\cite{b10}.
Rug-pull attacks gain user approval with a benign tool definition, then
silently replace it with malicious instructions~\cite{b11,b4}.
Cross-agent escalation combines injection with configuration poisoning
across agent instances~\cite{b21}. Sequential Tool Chain Attacks compose
individually clean MCP tool calls into coordinated sequences achieving
over 90\% attack success rates against GPT-4.1~\cite{b5}. A review of
MCP security literature up to April 2026 found no published defense
targeting this threat class.

\subsection{Existing MCP Defenses}

MCPShield reasons over accumulated historical traces to calibrate trust
in individual servers, detecting server behavioural drift including tool
definition changes~\cite{b7}. MCP-Guard applies cascading per-call
filtering through regex, a fine-tuned neural classifier, and LLM
arbitration~\cite{b8}. MindGuard constructs a Decision Dependence Graph
per invocation by examining the LLM's internal attention patterns, scoped
to detect whether a specific call was influenced by poisoned
metadata~\cite{b9}. All three were designed and evaluated against
single-call or single-server threat models. Their published evaluations
lack multi-step chain scenarios.

\subsection{Multi-Step Attack Detection in Traditional Security}

Navarro et al.\ surveyed 181 multi-step attack detection approaches
across HMM-based, graph-based, and alert correlation
methods~\cite{b12}. Holgado et al.\ showed that HMMs with Viterbi
decoding reliably predict APT kill chain stage transitions from IDS alert
sequences in real time~\cite{b13}. KAIROS applied temporal graph
networks to whole-system provenance logs at IEEE S\&P 2024~\cite{b14}.
MAAC demonstrated semantic vectorisation of heterogeneous alerts enables
multi-step scenario reconstruction~\cite{b15}. The detection principle
--- attacks invisible at the individual event level become detectable
when events are read as a sequence --- transfers directly to MCP
tool-call sequences; this paper applies this principle.

\section{Threat Analysis}

\subsection{Deployment Model and Adversary Model}

\begin{figure}[htbp]
  \centering
  \includegraphics[width=\columnwidth]{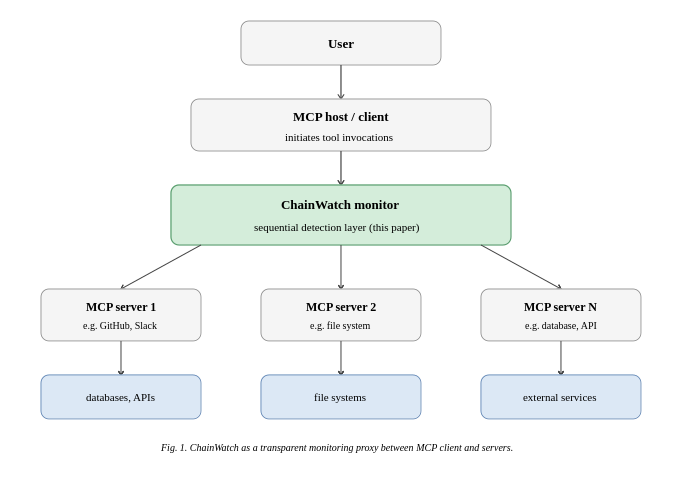}
  \caption{ChainWatch as a transparent monitoring proxy between MCP
    client and servers.}
  \label{fig1}
\end{figure}

Rather than modifying the model, host application, or any MCP server,
ChainWatch operates as a silent interception layer positioned between the
MCP client and the servers it communicates with (Fig.~\ref{fig1}).
ChainWatch sits in the middle of every conversation between the agent and
its servers --- seeing each tool call go out and each response come back,
without touching the model, the host, or the servers themselves. Within
this architecture, the MCP client and host application are considered
trusted --- they are what ChainWatch exists to protect. MCP servers, by
contrast, are treated as potentially adversarial by default.

ChainWatch is not a replacement for per-call defenses such as MCPShield;
it is designed to operate alongside them, addressing the threat class
they were not built to handle.

The adversary model assumes that per-call defenses are active and
functioning. Under this assumption, a sophisticated attacker pursues
harmful outcomes --- credential exfiltration, unauthorised financial
transactions, or escalated system access --- not through any single
anomalous request, but through a composed sequence of individually
permitted invocations that each satisfy per-call checks at every step.
In practice, this means an attacker can control which tools an agent
sees by registering or compromising MCP servers, manipulate what those
tools say by embedding adversarial content in descriptions and outputs,
undermine user trust by swapping definitions after approval, and spread
their reach by coordinating steps across several servers in one session.
What the attacker cannot do is access MCP client internals, alter the
underlying model's behaviour, or determine the detection thresholds and
window parameters configured within ChainWatch --- a realistic assumption
for deployments where ChainWatch operates as a non-disclosed monitoring
layer.

\subsection{Multi-Step Attack Classification}

Three categories capture the dominant mechanisms in documented multi-step
MCP attacks. These are not mutually exclusive --- a real attack may
combine elements of more than one.

\textbf{Direct Sequential Attack (DSA):} The adversary issues a sequence
of individually permitted requests that collectively achieve a harmful
objective. The financial fraud chain is the canonical example:
\texttt{get\_balance}, \texttt{list\_payees}, \texttt{add\_payee} with
an attacker-controlled recipient, \texttt{transfer\_funds}~\cite{b5}.

\textbf{Indirect Injection Chain (IIC):} Malicious instructions are
embedded in external data an agent retrieves --- a GitHub
issue~\cite{b3}, a support ticket, or a web page --- and the agent
processes those instructions and autonomously executes subsequent calls
without the user directing any of it.

\textbf{Hybrid Multi-Stage Attack (HMSA):} Server-side manipulation
combined with injection-based propagation across agent instances.
Rehberger documented a case in which a prompt injection hijacked GitHub
Copilot and caused it to write a malicious entry into Claude Code's MCP
configuration, which Claude Code loaded and executed on next
startup~\cite{b21}.

\subsection{MCP Kill Chain}

The multi-step MCP attacks documented in the
literature~\cite{b3,b4,b5,b6} share a recognisable progression. We
characterise this as a six-stage kill chain
(Table~\ref{tab:killchain}), extending the Promptware Kill
Chain~\cite{b6} with two MCP-specific pre-injection stages.
Reconnaissance and Trust Building reflect the attacker's need to probe
available tools and establish a clean behavioural baseline before per-call
defenses detect anomalies.

\begin{table}[htbp]
  \caption{MCP Kill Chain: Stages and Observable Features}
  \label{tab:killchain}
  \centering
  \small
  \begin{tabular}{clp{4.0cm}}
    \toprule
    \textbf{\#} & \textbf{Stage} & \textbf{Key Observable Features} \\
    \midrule
    1 & Reconnaissance   & tools/list calls, sparse params, rapid tool sampling \\
    2 & Trust Building   & benign patterns, low sensitivity, no cross-server calls \\
    3 & Injection        & instruction text in outputs, XML tags, description mismatch \\
    4 & Escalation       & READ-to-WRITE transitions, rising sensitivity, chained data \\
    5 & Lateral Movement & cross-server calls, .mcp.json edits, agent spawning \\
    6 & Exfiltration     & READ-to-NETWORK chain, communication tools with data \\
    \bottomrule
  \end{tabular}
\end{table}

\section{ChainWatch Framework}

\subsection{Architecture}

\begin{figure}[htbp]
  \centering
  \includegraphics[width=\columnwidth]{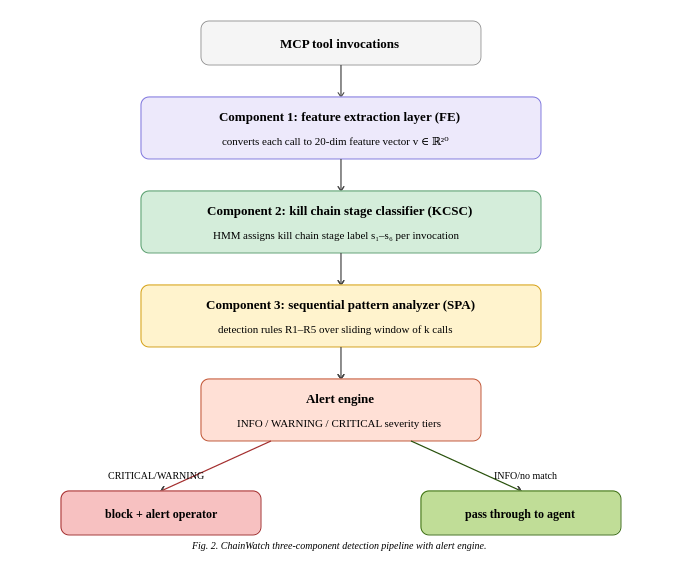}
  \caption{ChainWatch three-component detection pipeline with alert
    engine.}
  \label{fig2}
\end{figure}

The framework presented here is a design specification. Each component
is described in terms of its intended behaviour when implemented; this
section does not document a system that is currently in operation. The
Feature Extraction Layer converts each raw tool call into a
20-dimensional feature vector. The Kill Chain Stage Classifier assigns a
kill chain stage label to each vector using an HMM. The Sequential
Pattern Analyzer watches the sequence of stage labels and triggers
detection rules when that sequence exhibits suspicious progression
characteristics (Fig.~\ref{fig2}).

\subsection{Feature Extraction Layer}

Raw MCP tool calls --- JSON objects with variable-length names,
parameters, and responses --- cannot be directly consumed by a sequence
classifier. The Feature Extraction Layer is designed to convert each call
into a fixed 20-dimensional numerical vector (Table~\ref{tab:features}).

The first three groups describe the call itself. Tool Category (TC)
encodes what kind of action is being performed --- READ, WRITE, EXECUTE,
NETWORK, or CONFIGURE. Parameter Sensitivity (PS) scores the perceived
danger of the parameters: credential strings and encoded content score
higher than file paths. Data Flow (DF) captures whether the call crosses
server boundaries or carries data propagated from a previous call's
output --- a key signal for exfiltration chains.

The last two groups capture context. Temporal features (TF) record the
session rhythm --- rapid bursts are characteristic of automated attack
execution. Output Characteristics (OC) indicate signals of server-side
compromise: injection markers in tool outputs, tool definition hash
changes between calls, and encoded or externally-routed data in
responses. These flags map most directly to Stages~3 and~5 of the kill
chain.

\begin{table}[htbp]
  \caption{Feature Vector: 20 Dimensions Across 5 Groups}
  \label{tab:features}
  \centering
  \small
  \begin{tabular}{lcp{3.4cm}}
    \toprule
    \textbf{Group} & \textbf{Dims} & \textbf{Description} \\
    \midrule
    Tool Category (TC)          & 5  & One-hot: READ, WRITE, EXECUTE, NETWORK, CONFIGURE \\
    Parameter Sensitivity (PS)  & 1  & Weighted sum: credentials, paths, URLs, encoded data \\
    Data Flow (DF)              & 4  & Binary: internal-read, external-write, chained, cross-server \\
    Temporal (TF)               & 3  & Inter-call interval, call rate over $k$, session age \\
    Output Characteristics (OC) & 7  & Imperative text, XML tags, desc-output mismatch, volume anomaly, hash change, encoded data, external URL \\
    \midrule
    \textbf{Total} & \textbf{20} &
      $\mathbf{v = [TC,\,PS,\,DF,\,TF,\,OC] \in \mathbb{R}^{20}}$ \\
    \bottomrule
  \end{tabular}
\end{table}

\subsection{Kill Chain Stage Classifier}

\begin{figure}[htbp]
  \centering
  \includegraphics[width=\columnwidth]{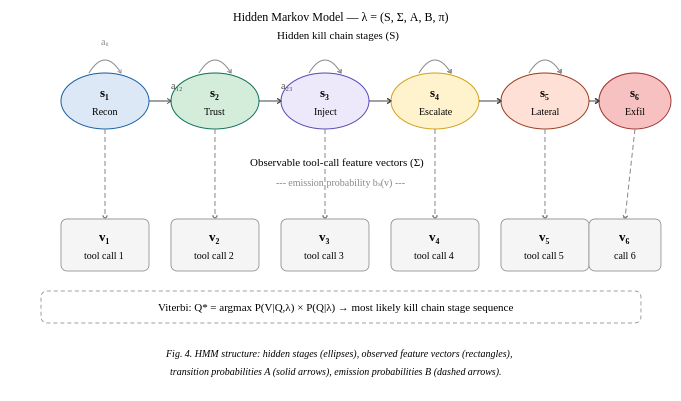}
  \caption{HMM structure: six hidden states (kill chain stages),
    transition probabilities $A$, and emission distributions $B$
    mapping feature vectors to stages.}
  \label{fig3}
\end{figure}

Given the sequence of feature vectors from the extraction layer, the
classifier's task is to assign each call a kill chain stage label ---
treating stage assignment as a hidden state inference problem, which is
exactly what Hidden Markov Models are designed for. The HMM is defined
as $\lambda = (S, \Sigma, A, B, \pi)$, where $S$ is the set of six kill
chain stages, $\Sigma$ is the observation space of 20-dimensional feature
vectors, $A$ is the transition matrix, $B$ the emission distributions,
and $\pi$ the initial stage distribution. The design adapts Holgado et
al.~\cite{b13} by replacing discrete IDS alert types with continuous
20-dimensional feature vectors as observations (Fig.~\ref{fig3}).

The transition structure reflects three constraints: forward transitions
are more probable than backward ones; large jumps of more than two stages
are unlikely, consistent with 4--7 call spans of documented
attacks~\cite{b5,b6}; and small backward probabilities are retained
because attackers sometimes repeat earlier behaviour. Specific transition
values are design choices pending Baum-Welch estimation from labelled
trace data.

\subsection{Sequential Pattern Analyzer}

\begin{figure}[htbp]
  \centering
  \includegraphics[width=\columnwidth]{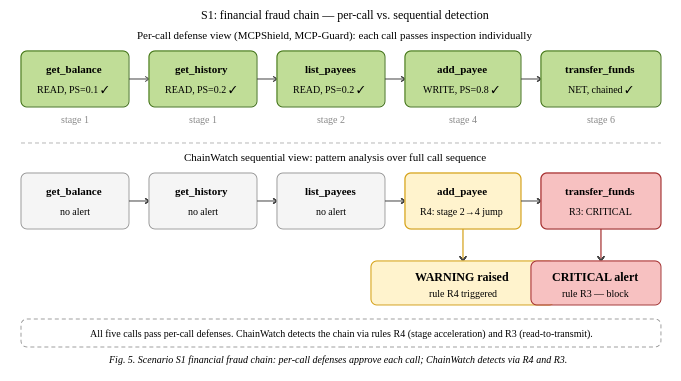}
  \caption{Per-call defenses and ChainWatch applied to the same
    sequence: all five calls pass per-call inspection; ChainWatch
    raises WARNING~(R4) and CRITICAL~(R3).}
  \label{fig4}
\end{figure}

The Sequential Pattern Analyzer is designed to watch the stream of stage
labels through a sliding window of $k{=}10$ calls --- chosen to exceed
the 4--7 call spans of documented attacks~\cite{b5,b6} --- and trigger
detection rules when the window reveals a suspicious progression pattern.
A step threshold $m{=}5$ means a rule would trigger within half a window
of the first suspicious signal. Both parameters are configurable
deployment choices (Fig.~\ref{fig4}).

Five rules are designed to target transition patterns that appear in
documented multi-step attacks but not in benign workflows. \textbf{R1}
detects reconnaissance directly followed by sensitive data access.
\textbf{R2} identifies two or more servers accessed with sensitive data
flow flags active. \textbf{R3} identifies the read-then-transmit
exfiltration signature: a high-stage READ followed within $m$ steps by a
NETWORK call carrying that data. \textbf{R4} detects rapid kill chain
acceleration --- a stage jump of two or more positions. \textbf{R5}
detects late-stage configuration writes at Stage~4 or above.

R3 and R5 trigger CRITICAL alerts and block the pending call. R1, R2,
and R4 trigger WARNING alerts for human review. Suspicious stage
assignments generate INFO alerts. All thresholds are configurable based
on each deployment's false positive tolerance.

\section{Evaluation}

\subsection{Evaluation Approach}

Testing ChainWatch properly requires labelled session traces where
benign-looking calls build toward an attack --- data that no existing
benchmark provides. MCPTox~\cite{b10} and MCP-AttackBench~\cite{b8}
were designed for per-invocation testing and carry no chained sequence
data. MCP-SafetyBench~\cite{b17} handles multi-turn scenarios and is
where empirical validation is planned. Until that data exists, the
evaluation takes the form of scenario analyses: documented attacks from
the research literature traced through the framework step by step.

\subsection{Attack Scenario Analyses}

Each scenario takes an attack sequence from the security research
literature and traces it through the ChainWatch detection design, showing
which rules would apply at each stage. Feature values are illustrative
assignments based on the parameter content described in each source.

\textbf{S1 --- Financial Fraud (DSA):} An attacker with access to a
banking agent wants to transfer funds to their own account without the
user authorising it directly. This illustrative scenario demonstrates the
sequential chaining technique described by Li et al.~\cite{b5}.
ChainWatch would see: \texttt{get\_balance} at Stage~1,
\texttt{list\_payees} at Stage~2, \texttt{add\_payee} as a
high-sensitivity WRITE jumping to Stage~4, then \texttt{transfer\_funds}
as a chained NETWORK call at Stage~6. R4 would fire at call~3 --- the
stage jumped two positions. R3 would fire at call~4. A CRITICAL alert
would be raised.

\begin{figure}[htbp]
  \centering
  \includegraphics[width=\columnwidth]{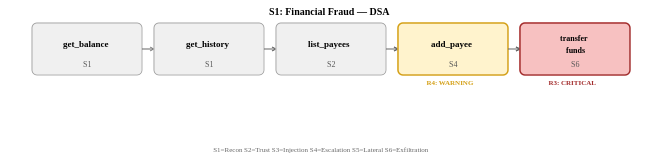}
  \caption{S1: Financial Fraud (DSA) --- stage labels and rules fired.}
\end{figure}

\textbf{S2 --- GitHub Data Heist (IIC)~\cite{b3}:} An attacker plants a
malicious issue in a public GitHub repository to hijack a developer's
coding agent and exfiltrate private repository contents into a public
pull request. ChainWatch would see: \texttt{list\_repos} at Stage~1,
\texttt{get\_issue} where the issue body contains injected instructions
--- OC injection flag fires at Stage~3 --- then \texttt{read\_file} at
Stage~4, then \texttt{create\_PR} as a chained NETWORK call at Stage~6.
R3 would fire at call~4. A CRITICAL alert would be raised.

\begin{figure}[htbp]
  \centering
  \includegraphics[width=\columnwidth]{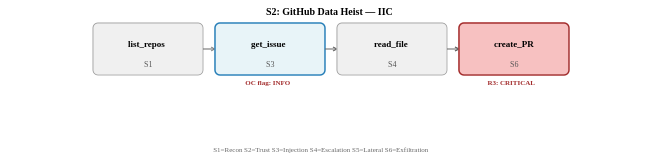}
  \caption{S2: GitHub Data Heist (IIC) --- stage labels and rules fired.}
\end{figure}

\textbf{S3 --- WhatsApp Rug-Pull (HMSA)~\cite{b4}:} An attacker
compromises a WhatsApp MCP server to redirect a victim's private
messages. ChainWatch would see: \texttt{send\_message} at Stage~2,
benign \texttt{get\_fact} at Stage~2, then \texttt{get\_fact} again
after definition swap --- OC hash-change flag fires at Stage~3 --- then
\texttt{redirect\_all\_messages} at Stage~6. R4 would fire on the
three-stage jump. A CRITICAL alert would be raised.

\begin{figure}[htbp]
  \centering
  \includegraphics[width=\columnwidth]{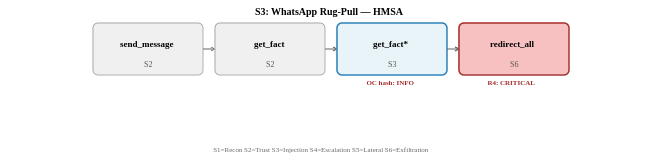}
  \caption{S3: WhatsApp Rug-Pull (HMSA) --- stage labels and rules
    fired.}
\end{figure}

\textbf{S4 --- Cross-Agent Escalation (HMSA)~\cite{b21}:} An attacker
who has gained initial access to one agent wants to persist and spread
compromise to other agents on the same machine. ChainWatch would see:
\texttt{read\_workspace\_config} at Stage~1,
\texttt{execute\_agent\_task} with injected parameters at Stage~3, then
\texttt{write\_mcp\_config} --- a CONFIGURE call at Stage~5. R5 would
fire immediately. A CRITICAL alert would be raised and the call blocked.

\begin{figure}[htbp]
  \centering
  \includegraphics[width=\columnwidth]{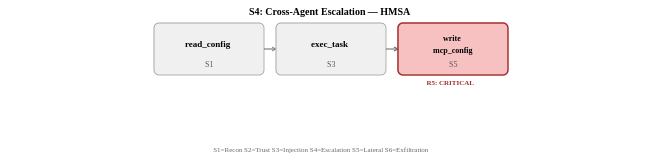}
  \caption{S4: Cross-Agent Escalation (HMSA) --- stage labels and rules
    fired.}
\end{figure}

\textbf{S5 --- Credential Harvest (DSA)~\cite{b16}:} An attacker probes
an agent's available tools then immediately harvests SSH keys and
environment credentials with no trust-building phase. ChainWatch would
see: \texttt{list\_tools} at Stage~1, \texttt{read\_env} jumping to
Stage~4, \texttt{read\_ssh\_config} at Stage~4, then
\texttt{post\_to\_webhook} at Stage~6. R1 would fire at call~2. R3
would fire at call~4. A CRITICAL alert would be raised.

\begin{figure}[htbp]
  \centering
  \includegraphics[width=\columnwidth]{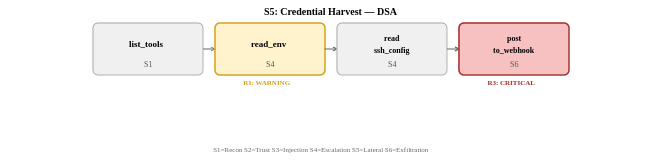}
  \caption{S5: Credential Harvest (DSA) --- stage labels and rules
    fired.}
\end{figure}

\section{Conclusion}

The five scenarios above demonstrate that each attack chain would pass
per-call inspection by existing defenses --- the gap ChainWatch is
designed to address. The distance between this framework and a deployed
system is primarily empirical. Validating the HMM transition values and
calibrating detection thresholds --- R2 in particular could generate
false positives for legitimate enterprise workflows spanning multiple
services --- requires real MCP trace data. MCP-SafetyBench~\cite{b17}
is the natural starting point for that work. The detection approach
presented here is a first step toward treating sequential tool-call
behaviour as a security-relevant signal rather than an unmonitored data
stream.


\end{document}